\documentclass[aps, pre, superscriptaddress]{revtex4-2} 
\usepackage[T1]{fontenc}

\usepackage{amsmath,fixmath}

\usepackage[usenames,dvipsnames]{xcolor}
\definecolor{shadecolor}{gray}{0.9}
\usepackage{lineno}

\usepackage{graphicx}
\usepackage[labelfont=bf,justification=raggedright]{caption}
\usepackage{subcaption}
\usepackage{booktabs}
\usepackage{tabularx}

\usepackage{algorithm}
\usepackage{algpseudocode}
\usepackage{listings}
\usepackage{fancyvrb}
\fvset{fontsize=\normalsize}

\usepackage{natbib}

\usepackage[colorlinks,linktoc=all]{hyperref}
\usepackage[all]{hypcap}
\hypersetup{citecolor=Violet}
\hypersetup{linkcolor=black}
\hypersetup{urlcolor=MidnightBlue}

\usepackage[nameinlink]{cleveref}

\usepackage{comment}

\newcommand{\be}{\begin{equation}}
\newcommand{\ee}{\end{equation}} 
\newcommand{\bea}{\begin{eqnarray}}
\newcommand{\eea}{\end{eqnarray}}

\newcommand{\grad}{{\nabla}} 

\renewcommand{\v}[1]{\mathbf{#1}} 
\renewcommand{\bf}[1]{\textbf{#1}} 

\newcommand{\rup}[1]{\left[#1\right]}

\renewcommand{\t}[1]{\tilde{#1}}

\renewcommand{\ref}[1]{[\ref{#1}]}
\crefname{equation}{Eq.}{Eqs.}
\crefname{section}{Sec.}{Secs.}
\crefname{figure}{Fig.}{Figs.}

\usepackage{titlesec}

\titlespacing\section{0pt}{12pt plus 4pt minus 2pt}{8pt plus 2pt minus 2pt}

\usepackage{amsfonts}
\usepackage{amssymb}
\usepackage{amsmath}
\usepackage{dsfont}

\usepackage[normalem]{ulem} 

\begin{document}
\title{Convergence properties of optimal transport-based temporal networks}
\author{Diego Baptista}
\affiliation{ Max Planck Institute for Intelligent Systems, Cyber Valley, Tuebingen, 72076, Germany}
\author{Caterina De Bacco}
\affiliation{ Max Planck Institute for Intelligent Systems, Cyber Valley, Tuebingen, 72076, Germany}

\begin{abstract}
We study network properties of networks evolving in time based on optimal transport principles. These  evolve from a structure covering uniformly a continuous space towards an optimal design in terms of optimal transport theory. At convergence, the networks should optimize the way resources are transported through it.
As the network structure shapes in time towards optimality, its topological properties also change with it. The question is how do these change as we reach optimality. We study the behavior of various network properties 
on a number of network sequences evolving towards optimal design and find that the transport cost function converges earlier than network properties and that these monotonically decrease. This suggests a mechanism for designing optimal networks by compressing dense structures. We find a similar behavior in networks extracted from real images of the networks designed by the body shape of a slime mold evolving in time.   
\end{abstract}
\pacs{}

\maketitle 
\section{Introduction}


Optimal Transport (OT) theory studies optimal ways of transporting resources in space \cite{villaniot,santambrogio2015optimal}.
The solutions are optimal paths that connect  sources to sinks (or origins to destinations) and the amount of flow traveling through them. 
In a general setting one may start from a continuous space in 2D, arbitrarily set sources and sinks, and then look for such optimal paths without any pre-defined underlying topology. Empirically, in many settings, these paths resemble network-like structures that embed optimality, in that traffic flowing along them is minimizing a transport cost function. Among the various ways to compute these solutions \cite{peyre2019computational}, a promising and computationally efficient recent approach is that of Facca et al. \cite{facca2016towards,facca2019numerics,facca2020branch}, which is based on solving a set of equations (the so-called \textit{Dynamical Monge-Kantorovich} (DMK) equations). This starts from an initial guess of the optimal paths that are then updated in time until reaching a steady state configuration. At each time step, one can automatically extract a principled network from a network-like structure using the algorithm proposed in Baptista et al. \cite{baptista2020network}. This in turn allows observing a sequence of network structures that evolves in time towards optimality, as the dynamical equations are iterated. While we know that the transport cost function is decreasing along this trajectory, we do not know how network properties on these structures evolve. For instance, in terms of the total number of edges or nodes, one may intuitively expect a monotonically decreasing behavior, from a topology  covering uniformly the whole space, towards a compressed one only covering a subset of it efficiently. 
Analyzing the properties of networks that provide optimal transport efficiency is relevant in many contexts and has been explored in several works \cite{corson2010fluctuations,bohn2007structure,durand2006architecture,katifori2010damage}. However, these studies usually consider pre-existing underlying topologies that need to be optimized. Moreover, they focus on network properties at convergence. Here instead we consider the situation where a network can be designed in a continuous 2D space, i.e. with no pre-defined underlying topology, and monitor the whole evolution of network properties, in particular away from convergence. While this question has been explored in certain biological networks \cite{baumgarten2013functional,baumgarten2010plasmodial,dirnberger2017characterizing,westendorf2016quantitative}, a systematic investigation of this intuition is still missing. 
In this work, we address this problem by considering several optimization settings, extracting their optimal networks, and then measuring core network properties on them. We find that network sequences show similar convergence patterns of those exhibited by their continuous counterparts. However, topological features of optimal networks tend to develop slightly slower than total cost function minimization. We also find that, in some cases, this delay in convergence presented by the networks might give better representations than those extracted at other cost-based convergence times. Finally, we analyze real data of the \textit{P. polycephalum} slime mold evolving its network-like body shape in time as it explores the space foraging. We use networks extracted from images generated in wet-lab experiments \cite{dirnberger2017introducing}, and analyze their topological features. Pattern matches can be found between synthetic graphs and this family of real networks. \\
Understanding how network topology evolves towards optimality may shed light on broader questions about optimal network properties and how to obtain them.

\begin{figure}[!ht]
    \centering
\begin{subfigure}[b]{1\textwidth}
\includegraphics[width=\textwidth]{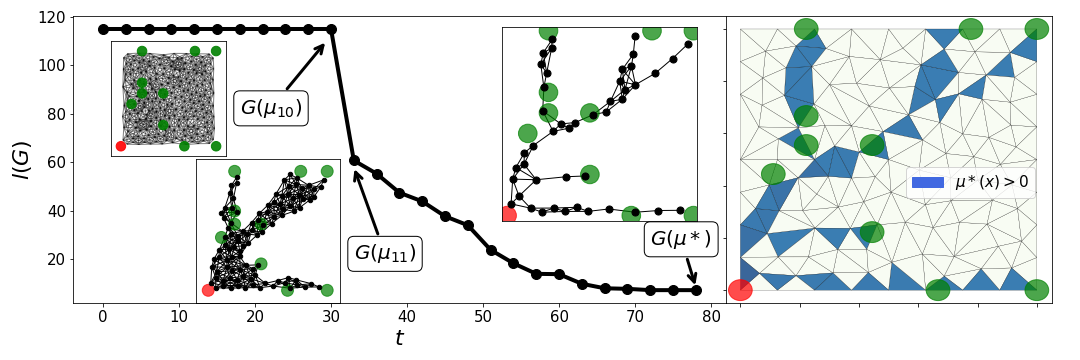}
\end{subfigure} 
\caption{\textbf{Temporal networks.}  On the left, the total length (i.e. sum of the edge lengths, as a function of time $t$; the networks inside the insets correspond to different time steps.  On the right, optimal transport density $\mu^*$; triangles are a $[0,1]^2$ discretization. In both plots, red and green circles correspond to the support of $f^+$ and $f^-$, i.e. sources and sinks, respectively. This sequence is obtained for $\beta = 1.5$.}\label{fig:image1}
\end{figure}

\section{The model}\label{sec:model}


\paragraph{The Dynamical-Monge Kantorovich set of equations.} We now present the main ideas of how to extract sequences of networks that converge towards an optimal configuration, according to optimal transport theory. We start by introducing the dynamical system of equations regulating this, as proposed by Facca et al. \cite{facca2016towards,facca2019numerics,facca2020branch}.  We assume that the problem is set on a continuous 2-dimensional space $\Omega \in \mathbb{R}^{2}$, i.e. there is no pre-defined underlying network structure. Instead, one can explore the whole space to design an optimal network topology, determined by a set of nodes and edges, and the amount of flow passing through each edge. Sources and sinks of a certain mass (e.g. passengers in a transportation network, water in a water distribution network) are displaced on it. We denote these with a ``forcing'' function $f(x)=f^+(x)-f^-(x)\in \mathbb{R}$,  describing the flow generating sources $f^+(x)$ and sinks $f^-(x)$ (also known as \textit{source} and \textit{target} distributions, respectively). It is assumed that $\int_\Omega f(x)dx  = 0$ to ensure mass balance. We suppose that the flow is governed by a transient Fick-Poiseuille type flux $q=- \mu \grad u$, where $\mu,u$ and $q$ are called \textit{conductivity} (or \textit{transport density}), \textit{transport potential} and \textit{flux}, respectively. Intuitively, the conductivity can be seen as proportional to the size of the edges where mass can flow, the potential could be seen as pressure on nodes, thus determining the flux passing on it.

The set of \textit{Dynamical Monge-Kantorovich} (DMK) equations is given by:
\bea\label{eqn:ddmk}
-\nabla \cdot (\mu(t,x)\nabla u(t,x)) &=& f^+(x)-f^-(x) \,, \label{eqn:ddmk1}\\
\frac{\partial \mu(t,x)}{\partial t}  &=& \rup{\mu(t,x)\nabla u(t,x)}^{\beta} - \mu(t,x) \,, \label{eqn:ddmk2}\\
\mu(0,x)  &=& \mu_0(x) > 0  \label{eqn:ddmk3} \,,
\eea
where $\nabla=\nabla_{x}$. Eq. (\ref{eqn:ddmk1}) states the spatial balance of the Fick-Poiseuille flux and is complemented by no-flow Neumann boundary conditions; Eq. (\ref{eqn:ddmk2}) enforces the dynamics of this system and Eq. (\ref{eqn:ddmk3}) is the initial configuration, this can be thought of as an initial guess of the solution. The parameter $\beta$ (traffic rate) tunes between various optimization setting: for $\beta<1$ we have congested transportation where traffic is minimized, $\beta>1$ is branched transportation where traffic is encouraged to consolidate along fewer edges, and $\beta=1$ is shortest path-like. In this work we only consider the branched transportation regime $1<\beta<2$, as this is the only one where meaningful network structures can be extracted \cite{baptista2020network}. 

Solutions $(\mu^*, u^*)$ of Eqs. (\ref{eqn:ddmk1})-(\ref{eqn:ddmk3}) minimize the transportantio n cost function $\mathcal{L}(\mu,u)$ (\cite{facca2016towards,facca2019numerics,facca2020branch}), defined as:
\bea\label{eqn:L}
& \mathcal{L}(\mu,u) := \mathcal{E}(\mu,u)+ \mathcal{M}(\mu,u) \\
& \mathcal{E}(\mu,u) := \dfrac{1}{2}\int_{\Omega} \mu |\grad u|^2 dx, \  \ \mathcal{M}(\mu,u) :=  \dfrac{1}{2}\int_{\Omega} \dfrac{\mu^{\frac{(2-\beta)}{\beta}}}{2-\beta} dx \quad.
\eea

$\mathcal{L}$ can be thought of as a combination of $\mathcal{M}$, the total energy dissipated during the transport (or network operating cost) and $\mathcal{E}$, the cost to build the network infrastructure (or infrastructural cost).

\subsection{Network sequences}


The conductivity $\mu$  at convergence  regulates where the mass should travel for optimal transportation. This is a function of a 2-dimensional space, it can be turned into a principled network $G(\mu)$ (a set of nodes, edges, and weights on them) by using the method proposed by \cite{baptista2020network}, which in turn determines the design of the optimal network. While the authors  of that work considered only values at convergence, this method is still valid at any time step, in particular at time steps before convergence. This then leads to a sequence of networks evolving in time as the DMK equations are iterated. Fig. \ref{fig:image1} shows three networks built using this method at different time steps. The leftmost inset is the densest representation that one can build from the shown discretization of the space (a triangulation), as in the plot on the right side: all the nodes are connected to all of their closest neighbors. This is what happens at initial time steps where the network is built from mass uniformly displaced in space, as per uniform initial condition. On the other hand, the rightmost network is built from  a $\mu$ at convergence, consolidated on a more branched structure.

Formally, let $\mu(x,t)$ be a \textit{transport density} (or \textit{conductivity}) function of both time and space obtained as a solution of the DMK model. We denote it as the sequence $\{\mu_t\}_{t=0}^T$, for some index $T$ (usually taken to be that of the convergent state). Every $\mu_{t}$ is the $t$-th update of our initial guess $\mu_0$, computed by following the rules described in Eqs. (\ref{eqn:ddmk1})-(\ref{eqn:ddmk3}). This determines a sequence of networks $\{ G(\mu_t)\}_{t=0}^T$ extracted from $\{\mu_t\}_{t=0}^T$ with \cite{baptista2020network}. Fig. \ref{fig:image1} shows three different snapshots of one of the studied sequences.

\paragraph{Convergence criteria.} Numerical convergence of the DMK equations (\ref{eqn:ddmk1})-(\ref{eqn:ddmk3}) can be arbitrarily defined. Typically, this is done by fixing a threshold $\tau$, and stopping the algorithm when the cost does not change more than that between successive time steps. However, when this threshold is too small ($\tau=10^{-12}$ in our experiments), the cost or the network structure may consolidate to a constant value way in advance, compared to the algorithmic one. Thus, to meaningfully establish when is network optimality reached, we consider as convergence time  the first time step when the transport cost, or a given network property, reaches a value that is smaller or equal to a certain fraction $p$ of the value reached by the same quantity at algorithmic convergence (in the experiments here we use $p=1.05$). We refer to $t_\mathcal{L}$ and  $t_P$ for the convergence in times in terms cost function or a network property, respectively.



\paragraph{Network properties.} We analyze the following main network properties for the different networks in the sequences and for different sequences. Denote with $G$ one of the studied networks belonging to some sequence $\{ G(\mu_t)\}_{t=0}^T$. We study the following properties relevant to the design of networks for optimal transport of resources through it.

\begin{itemize}
\item $|N|$, total number of nodes;
\item $|E|$, total number of edges;
\item \textit{total length} $l(G)=\sum_e l(e)$, i.e. the sum of the lengths of every edge. Here $l(e)$ is the Euclidean distance between the nodes endpoints of $e$;
\item Average degree, the mean number of neighbors per node;
\item $bif(G)$, the number of bifurcations; a \textit{bifurcation} is a node with degree greater than 2;
\item $leav(G)$, the number of leaves; a leave is a node with degree equal to 1.
\end{itemize}

\section{Results on synthetic data}\label{sec:synthetic}
%
To study the behavior of network structures  towards optimality, we perform an extensive empirical analysis as follows. We generate synthetic data considering a set of optimal transport problems, determined by the configuration of sources and sinks. In fact, the final solutions strongly depend on how these are displaced in space. We consider here a scenario where we have one source and many sinks, which is a relevant situation in many applications. For instance, in biology, this would be the case for a slime mold placed on a point in space (the source) and looking for multiple sources of food (the sinks).
Formally, consider a set of points $S = \{s_0,s_1,...,s_M\}$  in the space $\Omega = [0,1]^2,$ and $0<r$ a positive number. We define the distributions $f^+$ and $f^-$ as 
$$
f^+(x) \propto  \mathds{1}_{R_0}(x), \ \  f^-(x) \propto  \sum_{i>0} \mathds{1}_{R_i}(x)
$$  

where $\mathds{1}_{R_i}(x) := 1,$ if $x\in R_i$, and $\mathds{1}_{R_i}(x) := 0$, otherwise; $R_i$ is the circle of center $s_i$ and radius $r$ (the value of $r$ is automatically selected by the solver based on the discretization of the space); and the proportionality is such that $f^+$ and $f^-$ are both probability distributions.  The transportation cost is that of Eq. (\ref{eqn:L}).

\paragraph{Data generation.} We generate 100 transportation problems by fixing the location of the source $s_0=(0,0)$ (i.e. the support of $f^+$ at $(0,0)$), and sampling 15 points $s_1,s_2,...,s_M$  uniformly at random from a regular grid (see supplementary information). By choosing points from vertices of a grid, we ensure that the different sinks are sufficiently far from each other, so they are not redundant. We start from an uniform, and thus non-informative, initial guess for the solution, $\mu_0(x) = 1, \forall x$. We fix the maximum number of iterations to be 300.  We say that the sequence $\{\mu_t\}_{t=0}^T$ \textit{converges} to a certain function $\mu^*$ at iteration $T$ if either $|\mu_T-\mu_{T-1} |<\tau,$  for a \textit{tolerance} $\tau\in (0,1],$ or $T= 300$. For the experiments reported in this manuscript, the tolerance $\tau$ is set to be $10^{-12}$. We consider different values of $\beta \in [1.1,1.9]$, thus exploring various cost functions within the branched transportation regime. Decreasing $\beta$ from 2 to 1 results in traffic being more penalized, or consolidation of paths into fewer highways less encouraged. In total, we obtain  900 network sequences, each of them containing between 50 and 80 networks.

\begin{figure}[!h] 
    \centering
\begin{subfigure}[b]{0.32\textwidth}
\includegraphics[width=\textwidth]{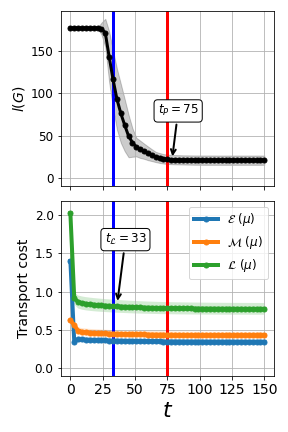} 
\end{subfigure}
\begin{subfigure}[b]{0.32\textwidth}
\includegraphics[width=\textwidth]{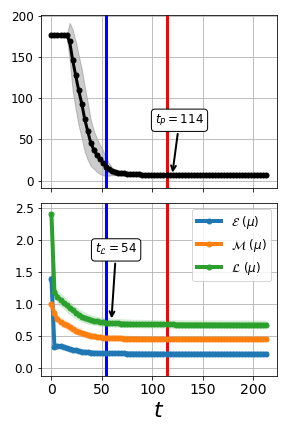}
\end{subfigure}
\begin{subfigure}[b]{0.32\textwidth}
\includegraphics[width=\textwidth]{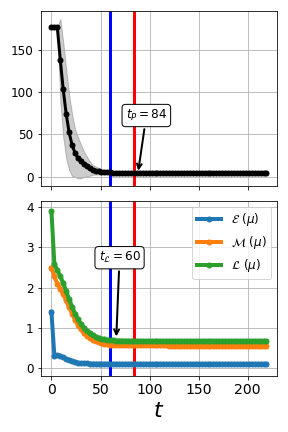}
\end{subfigure} 
\caption{\textbf{Total length and Lyapunov cost.} Mean (markers) and standard deviations (shades around the markers) of the total length (top plots) and  of the Lyapunov cost, energy dissipation $\mathcal{E}$ and structural cost $\mathcal{M}$ (bottom plots), as functions of time $t$. Means and standard deviations are computed on the set described in Section \ref{sec:synthetic}. From left to right: $\beta=1.2, 1.5$ and $1.8$. Red and blue lines denote $t_P$ and $t_\mathcal{L}$. }\label{fig:image2}
\end{figure}

\vspace{-1cm}
\paragraph{Convergence: transport cost vs network properties.}  

Fig. \ref{fig:image2} shows a comparison between network properties and the cost function minimized by the dynamics.

\begin{figure}[!h]
    \centering
\begin{subfigure}[b]{.8\textwidth}
\includegraphics[width=\textwidth]{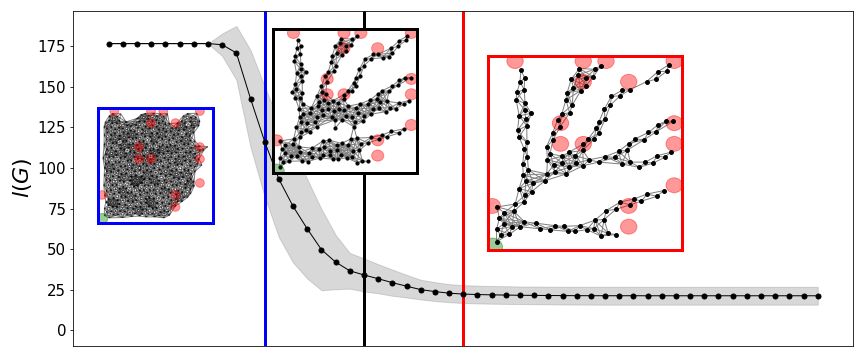}
\end{subfigure} 
\begin{subfigure}[b]{.8\textwidth}
\includegraphics[width=\textwidth]{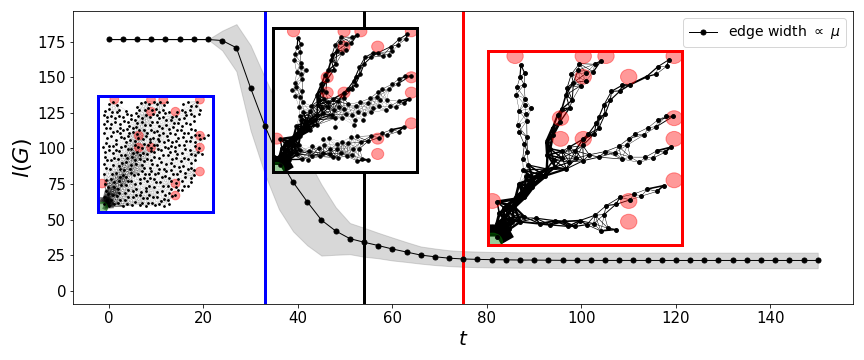}
\end{subfigure} 
\caption{\textbf{Network topologies for different convergence criteria.} Mean (markers) and standard deviations (shades around the markers) of total length $l(G)$ as a function of time. The red, black and blue vertical lines (and networks in the insets) correspond to $t_P,$ the average of $t_P$ and $t_\mathcal{L},$ and $t_\mathcal{L}$, respectively. Networks without (top row) and with (bottom) edge weights proportional to their $\mu$ are plotted at those times three time steps. Hence, the networks on top highlight the topological structure, while those on the bottom the flux passing through edges.  }\label{fig:image3}
\end{figure}

We observe that $t_P>t_\mathcal{L}$ in all the cases, i.e. convergence in the cost function is reached earlier than convergence of the topological property. Similar behaviors are seen for other values of $\beta\in[1.1,1.9]$ and for other network properties (see supplementary information). For smaller values of $\beta$ convergence in transport cost is reached faster, when the individual network properties are still significantly different from their value at convergence, see Fig. \ref{fig:image3} for an example in terms of total path length at $\beta=1.2$. In this case, while the cost function does not change much after $t_\mathcal{L}$, the network properties do instead. This may be because the solutions for $\beta$ close to 1 have non-zero $\mu$ on many edges but most of them have low values. Indeed, we find that the most important edges, measured by the magnitude of $\mu$ on them, are those corresponding to the topology found at a later time, when the network properties also converge, as shown in Fig. \ref{fig:image3} (bottom). This indicates that the dynamics first considers many edges, and distributes the fluxes optimally along fewer main roads. At the end, close to convergence, it focuses instead on removing redundant edges, those that have little flux traveling.

Finally, we notice how $t_\mathcal{L}$ is smaller for $\beta$ close to 1. This reflects the fact that in this case it is easier to find a solution to the optimization problem, as for increasing $\beta$ the configuration space gets roughed with many local optima \cite{facca2016towards,facca2019numerics,facca2020branch}.

\begin{figure}[!h]
    \centering
\begin{subfigure}[b]{0.3\textwidth}
\includegraphics[width=\textwidth]{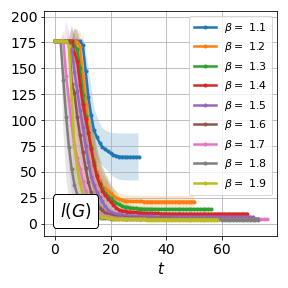}
\end{subfigure} 
        \begin{subfigure}[b]{0.3\textwidth}
        \includegraphics[width=\textwidth]{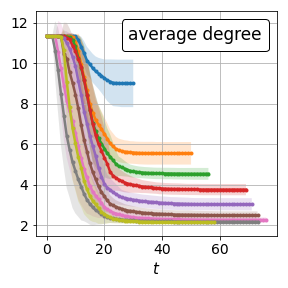}
    \end{subfigure} 
    \begin{subfigure}[b]{0.3\textwidth}
        \includegraphics[width=\textwidth]{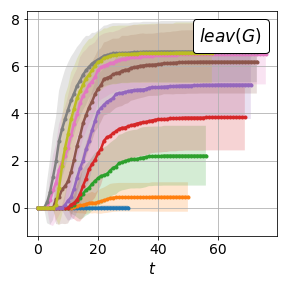}
    \end{subfigure}  
    \\
    \begin{subfigure}[b]{0.3\textwidth}
\includegraphics[width=\textwidth]{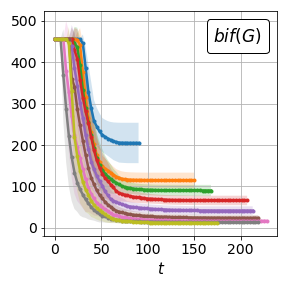}
\end{subfigure} 
        \begin{subfigure}[b]{0.3\textwidth}
        \includegraphics[width=\textwidth]{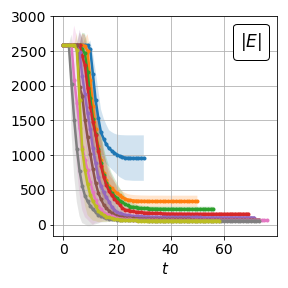}
    \end{subfigure} 
    \begin{subfigure}[b]{0.3\textwidth}
        \includegraphics[width=\textwidth]{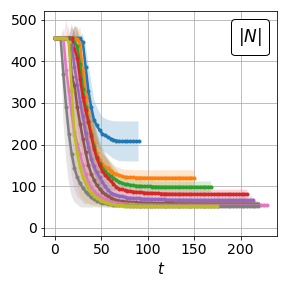}
    \end{subfigure} 
      
      \caption{\textbf{Evolution of network properties.} Mean (markers) and standard deviations (shades around the markers) of total length $l(G)$ (upper left), average degree (upper center), number of leaves $leav(G)$ (upper right), number of bifurcations $bif(G)$ (lower left), number of edges $|E|$ (lower center)and number of nodes $|N|$ (lower right), computed for different values of $\beta$ as a function of time.}\label{fig:image4}
\end{figure}

\begin{figure}[!ht]
    \centering
\begin{subfigure}[b]{1\textwidth}
\includegraphics[width=\textwidth]{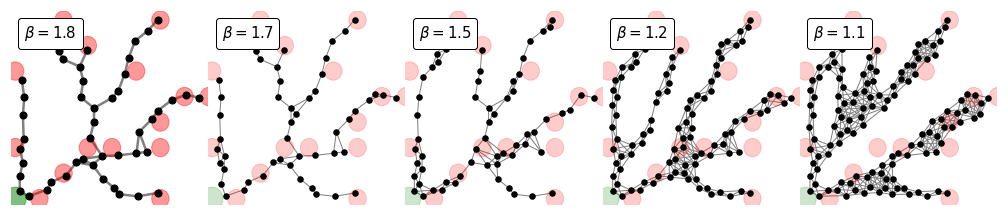}
\end{subfigure} 
\caption{\textbf{Example of optimal networks for various cost functions.} Networks extracted from the solutions of the same transportation problem but various $\beta$.  Green and red and  circles denote source and sinks.} \label{fig:image5}
\end{figure}

\paragraph{Convergence behavior of network properties.}  
Fig. \ref{fig:image4} shows how the various network properties change depending on the traffic rate. The plots show their mean values computed across times, for a fixed value of $\beta$. Notice that quantities like the total length, the average degree, the number of bifurcations, the number of edges and the number of nodes decrease in time, signalling that sequences reach steady minimum states. These are reached at different times, depending on $\beta$, with convergence reached faster for lower $\beta$. Moreover, mean values of these properties converge to decreasing values, as $\beta$ increases. This is explained by a cost function increasingly encouraging consolidations of paths on fewer edges. Finally, the magnitude of the gap between the different mean values of each property for different $\beta$ depends on the individual property. For instance, the average degree changes more noticeably between two consecutive values of $\beta$ than the total path length, which shows a big gap between the value at $\beta=1.1$ and all of the subsequent $\beta>1.1$, that have instead similar value of this property. This also shows that certain properties better reveal  the distinction between different optimal traffic regimes.  The number of leaves behaves more distinctly. In fact, it exhibits two different patterns: either it remains constantly equal to 0 ($\beta=1.1$) or it increases, and with different rates, as time gets larger ($\beta>1.1$). This number increases with $\beta$, as in this regime paths consolidate into fewer edges, thus leaving more opportunities for leaves.  To help intuition of the different optimal designs for various $\beta$, we plot the extracted networks at convergence in Fig. \ref{fig:image5}. The positions of source and sinks are the same in all cases. The network obtained for higher $\beta=1.8$ contains fewer edges and nodes than the others cases. On average, these networks have $bif(G)\approx13$ , $leav(G)\approx 7$,  $l(G)\approx 4$ and an average degree $\approx 2$.  These reveal various topological features on the converged networks of this traffic regime, that make it more distinct than others. For instance, having approximately 7 leaves implies that the dynamics builds networks with as many leaves as approximately half the number of sinks ($M=15$) in this transportation problem, while on the other hand, we can see that $bif(G)\approx M$, i.e., the number of bifurcations is almost as large as the number of sinks.

\subsection{Results on real networks of \textit{P. polycephalum} } 


In this section, we compare the properties of the sequences $ \{ G(\mu_t)\}_{t=0}^T$ to those extracted from real images of the slime mold \textit{P. polycephalum}. This organism has been shown to perform an optimization strategy similar to that modeled by the DMK equations of \Cref{sec:model}, while foraging for food in a 2D surface \cite{nakagaki2000maze,tero2007mathematical,tero2010rules}.   We extract these networks from images using the method proposed by \cite{baptista2020principled}. This pipeline takes as input a network-like image and uses the color intensities of its different pixels to build a graph, by connecting adjacent meaningful nodes. We choose 4 image sequences from the Slime Mold Graph Repository \cite{dirnberger2017introducing}. Every sequence is obtained by taking pictures of a \textit{P. polycephalum} placed in a rectangular Petri dish and following its evolution in time. Images are taken every 120 seconds from a fixed position. 

\begin{figure}[!ht]
    \centering
\begin{subfigure}[b]{1\textwidth}
\includegraphics[width=\textwidth]{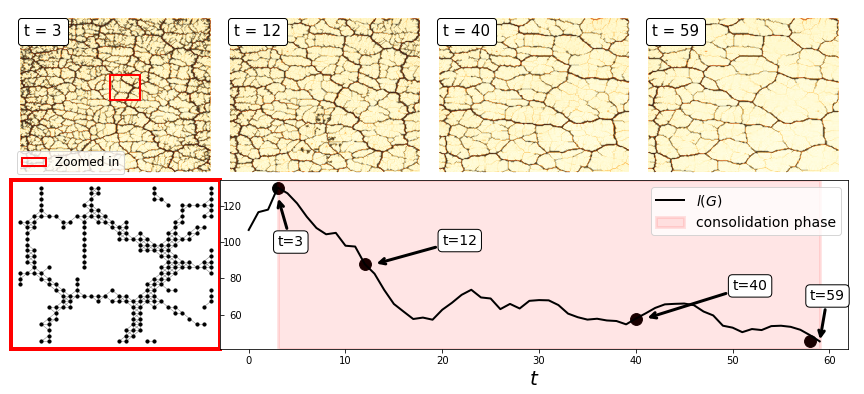}
\end{subfigure} 
\caption{\textbf{Network evolution of \textit{P. polycephalum.}} On top: \textit{P. polycephalum} images and networks extracted from them. Bottom left: a zoomed-in part of the graph shown inside the red rectangle on top. Bottom right: total length as a function of time. The red shade highlights a tentative consolidation phase towards optimality. } \label{fig:image6}
\end{figure}

We study the evolution of the total length  for every sequence. We show in Fig. \ref{fig:image6} the total length of the temporal network extracted from one of the mentioned image sequences (namely, image set \textit{motion12}; see supplementary information), together with different network snapshots.  As we can see from the lower rightmost plot,  the evolution of the total length of the extracted networks resembles that of the synthetic network sequences analyzed above.  This suggests that the DMK-generated sequences resemble the behavior of this real system in this time frame.   This could mean that the DMK dynamics realistically represents a consolidation phase towards optimality of real slime molds \cite{dirnberger2017introducing}. Similar results are obtained for other sequences (see supplementary information).

\section{Conclusions}

We studied  the  properties  of sequences of networks converging to optimal structures. 
Our results show that network sequences obtained from the solution of diverse transportation problems often minimize network properties at slower rates compared to the transport cost function. This suggests an interesting behavior of the DMK dynamics: first, it focuses on distributing paths into main roads while keeping many network edges. Then, once the main roads are chosen, it removes redundant ones where the traffic would be low. Measuring convergence of network properties would then reveal a more compressed network cleaned from redundant paths.
The insights obtained in this work may further improve our understanding of the mechanisms  governing  the design of optimal transport networks.

We studied here a particular set of transportation problems, one source, and multiple sinks. In this case, all the main network properties studied here show similar decaying behavior. However this analysis can be replicated for more complex settings, like multiple sources and multiple sinks \cite{lonardi2020optimal} or in multilayer networks \cite{ibrahim2021optimal}, as in urban transportation networks. Potentially, this may unveil different patterns of the evolution of the topological properties than those studied in this work.

Results on real networks suggest that the networks generated by the DMK dynamics (inspired by the \textit{P. polycephalum}) resemble realistic features. Strongly monotonic phases are not only typical of the mentioned slime molds but also a pattern in the artificially generated data. Alternative realistic behaviors may be seen by considering a modified version of the model described in Eq. (\ref{eqn:ddmk}) by adding non-stationary forcing terms. This may highlight a behavior different than the one observed in a consolidation phase, where a network converges to an optimal design and then does not change further. This is an interesting direction for future work. 

\paragraph{Acknowledgements}
The authors thank the International Max Planck Research School for Intelligent Systems (IMPRS-IS) for supporting Diego Baptista.

\bibliographystyle{splncs03}
\bibliography{references}

\newcommand{\beginsupplement}{%
        \setcounter{table}{0}
        \renewcommand{\thetable}{S\arabic{table}}%
        \setcounter{figure}{0}
        \renewcommand{\thefigure}{S\arabic{figure}}%
        \setcounter{equation}{0}
        \renewcommand{\theequation}{S\arabic{equation}}
         \setcounter{section}{0}
        \renewcommand{\thesection}{S\arabic{section}}
 }
 
\clearpage
\begin{widetext}

\section*{\centering Supplementary Information (SI)}

\section{Synthetic data}

\paragraph{Details of the studied transport problems.} As mentioned in the main manuscript, we consider a set of points $S = \{s_0,s_1,...,s_M\}$  in the space $\Omega = [0,1]^2,$ and $0<r$ a positive number, and we use this to define the distributions $f^+$ and $f^-$ as 
$$
f^+(x) \propto  \mathds{1}_{R_0}(x), \ \  f^-(x) \propto  \sum_{i>0} \mathds{1}_{R_i}(x)
$$  

where $R_i$ is the circle of center $s_i$ and radius $r$. The points $s_1,...,s_M$, the support of the sink, are sampled uniformly at random from a regular grid. The used grid and different realizations of the sampling are shown in Fig. \ref{fig:image7}. 

\begin{figure*}[hptb]
    \centering
\begin{subfigure}[b]{1\textwidth}
\includegraphics[width=\textwidth]{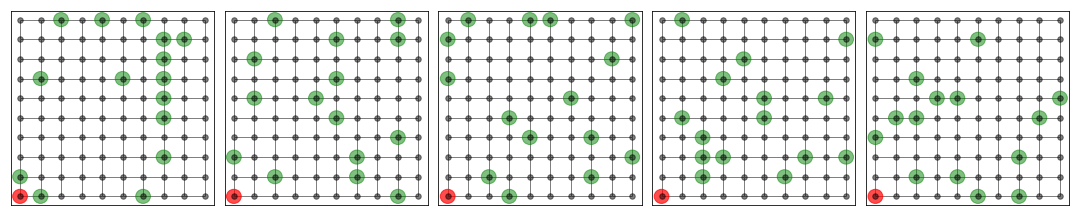}
\end{subfigure} 
\caption{\textbf{Support of $f^-$}. The nodes of the grid constitutes the set of candidates from which the support of $f^-$} \label{fig:image7}
\end{figure*}

\paragraph{Total length and Lyapunov cost.} We show in this section a figure like the one presented in the Fig. 2 of the main manuscript, for other values of $\beta.$ As mentioned in there, the properties show decreasing behaviors for which is always true that $t_P>t_\mathcal{L}$ (see Fig. \ref{fig:image8}).

\begin{figure*}[hptb]
    \centering
\begin{subfigure}[b]{0.32\textwidth}
\includegraphics[width=\textwidth]{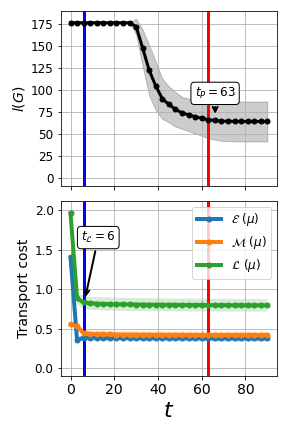}
\end{subfigure}
\begin{subfigure}[b]{0.32\textwidth}
\includegraphics[width=\textwidth]{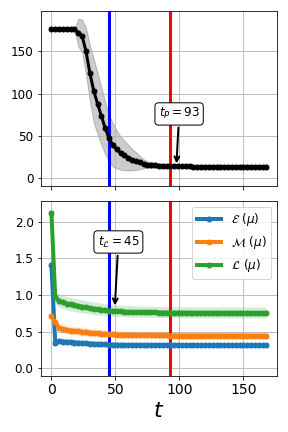}
\end{subfigure}
\begin{subfigure}[b]{0.32\textwidth}
\includegraphics[width=\textwidth]{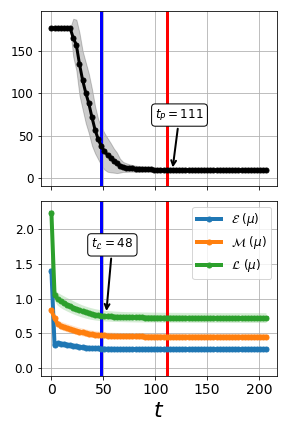}
\end{subfigure} 
\\
\begin{subfigure}[b]{0.32\textwidth}
\includegraphics[width=\textwidth]{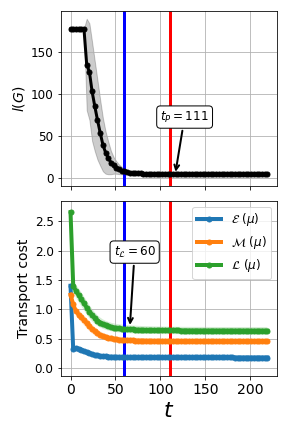}
\end{subfigure}
\begin{subfigure}[b]{0.32\textwidth}
\includegraphics[width=\textwidth]{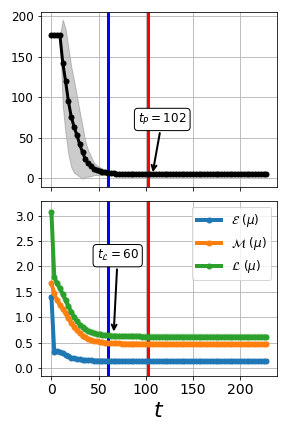}
\end{subfigure}
\begin{subfigure}[b]{0.32\textwidth}
\includegraphics[width=\textwidth]{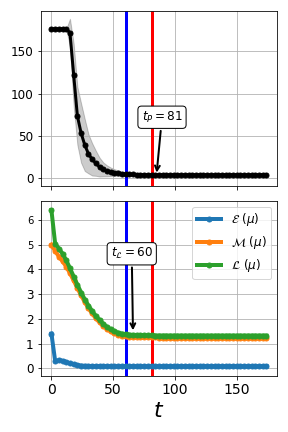}
\end{subfigure} 
\caption{\textbf{Total length and Lyapunov cost}. Top row: from left to right we see $\beta=1.1, 1.3$ and $1.4$. Bottom row: from left to right we see $\beta=1.6, 1.7$ and $1.9$. Mean and standard deviation of the total length $l(G)$ as function of time $t$; Bottom plot: Mean and standard deviation of  the Lyapunov cost $\mathcal{L}$, energy dissipation $\mathcal{E}$ and structural cost $\mathcal{M}$ of transport densities. Red and blue lines denote $t_P$ and $t_\mathcal{L}$ for $p = 1.05$.     }\label{fig:image8}
\end{figure*}

\paragraph{Network properties and Lyapunov cost.} We show in this section a figure like the one presented in the Fig. 2 of the main manuscript, for the other network properties (see Fig. \ref{fig:image9}).

\begin{figure*}[hptb]
    \centering
\begin{subfigure}[b]{0.32\textwidth}
\includegraphics[width=\textwidth]{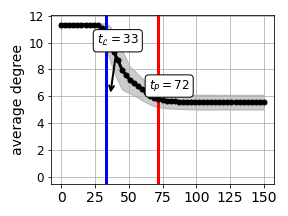}
\end{subfigure}
\begin{subfigure}[b]{0.32\textwidth}
\includegraphics[width=\textwidth]{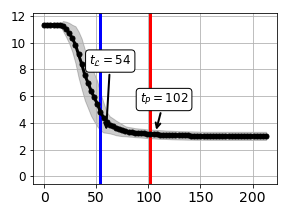}
\end{subfigure}
\begin{subfigure}[b]{0.32\textwidth}
\includegraphics[width=\textwidth]{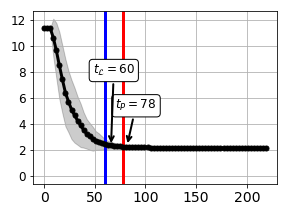}
\end{subfigure} 
\\
\begin{subfigure}[b]{0.32\textwidth}
\includegraphics[width=\textwidth]{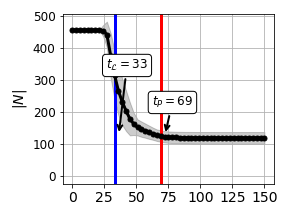}
\end{subfigure}
\begin{subfigure}[b]{0.32\textwidth}
\includegraphics[width=\textwidth]{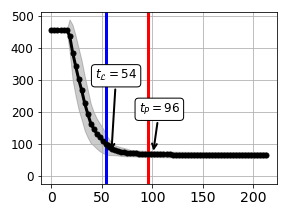}
\end{subfigure}
\begin{subfigure}[b]{0.32\textwidth}
\includegraphics[width=\textwidth]{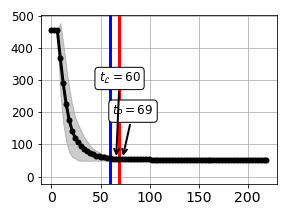}
\end{subfigure} 
\\
\begin{subfigure}[b]{0.32\textwidth}
\includegraphics[width=\textwidth]{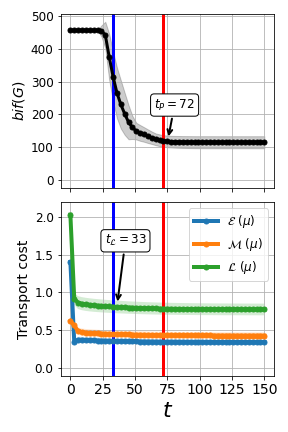}
\end{subfigure}
\begin{subfigure}[b]{0.32\textwidth}
\includegraphics[width=\textwidth]{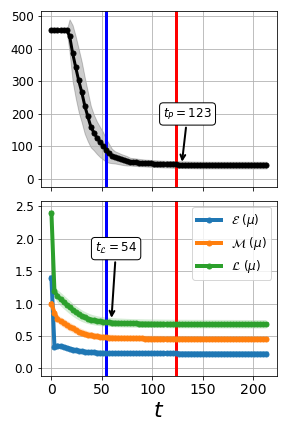}
\end{subfigure}
\begin{subfigure}[b]{0.32\textwidth}
    \includegraphics[width=\textwidth]{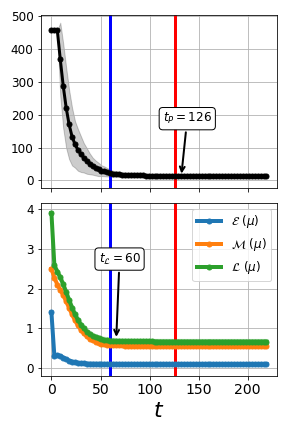}
\end{subfigure} 
\caption{\textbf{Other network properties and Lyapunov cost}. From left to right: $\beta=1.2, 1.5$ and $1.8.$ From top to bottom: Mean and standard deviation of the average degree, number of nodes $|N|,$ number of bifurcations $bif(G)$, and the Lyapunov cost $\mathcal{L}$, energy dissipation $\mathcal{E}$ and structural cost $\mathcal{M}$. Red and blue lines denote $t_P$ and $t_\mathcal{L}$ for $p = 1.05$.}\label{fig:image9}
\end{figure*}

\section{\textit{P. polycephalum} networks}

\paragraph{Data information.} In this section, we give further details about the used real data. As mentioned in the main manuscript, the images are taken from the Slime Mold Graph Repository \cite{dirnberger2017introducing}. The number of studied sequences $\{G_i\}_i^T$ equals 4. Every sequence's length $T$ changes depending on the amount of images provided in the repository, since different experiments need more o less shots. An experiment, as explained in the repository's documentation, consists of placing a slime mold inside a Petri dish with a thin sheet of agar and no sources of food. The idea, as explained by the creators, is to let the slime mold fully explore the Petri dish. Since the slime mold is initially lined up along one of the short side of the dish, the authors stop capturing images once the plasmodium is about to reach the other short side. 

\paragraph{Network extraction.} The studied network sequences are extracted from the image sets \textit{motion12, motion24, motion40} and \textit{motion79}, which are stored in the repository. Each image set contains a number of images ranging from 60 to 150, thus, obtained sequences exhibit diverse lengths.  Every network is extracted using the \textit{Img2net} algorithm described in \cite{baptista2020principled}. The main parameters of this algorithms are \texttt{N\_runs}, \texttt{t2}, \texttt{t3} and  \texttt{new\_size}. \texttt{N\_runs} controls how many times the algorithm needs to be run; \texttt{t2} (and \texttt{t3}) are the minimum value (and maximum) a pixel's grayscale value must be so it is considered as a node; \texttt{new\_size} is the size to which the input image must be downsampled before extracting the network from it. For all the experiments reported in this manuscript, the previously mentioned parameters are set to be \texttt{N\_runs}=1, \texttt{t2}= 0.25, \texttt{t3}=1 and  \texttt{new\_size}=180.

\paragraph{More network properties.} Other network properties are computed for the real systems referenced in this manuscript. Similar decreasing behaviors, like the one shown for the total length property in the main manuscript, are found for these properties; see Fig. \ref{fig:image10}. 

\begin{figure*}[hptb]
    \centering
\begin{subfigure}[b]{1\textwidth}
\includegraphics[width=\textwidth]{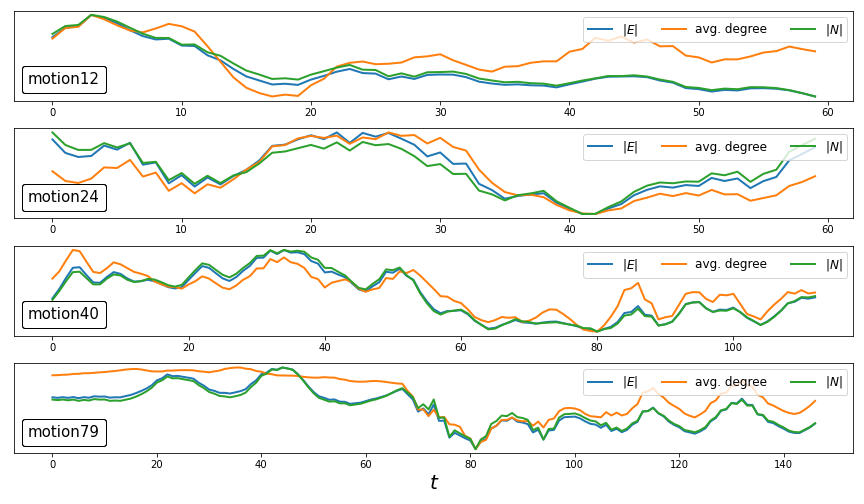}
\end{subfigure} 
\caption{\textbf{Network properties for \textit{P. polycephalum} sequences}. From top to bottom: \textit{motion12, motion24, motion40} and \textit{motion79}. Subfigures show the evolution of the properties $|E|,$ average degree and $|N|$ for every sequence as a function of time. } \label{fig:image10}
\end{figure*}

\begin{figure*}[hptb]
    \centering
\begin{subfigure}[b]{.8\textwidth}
\includegraphics[width=\textwidth]{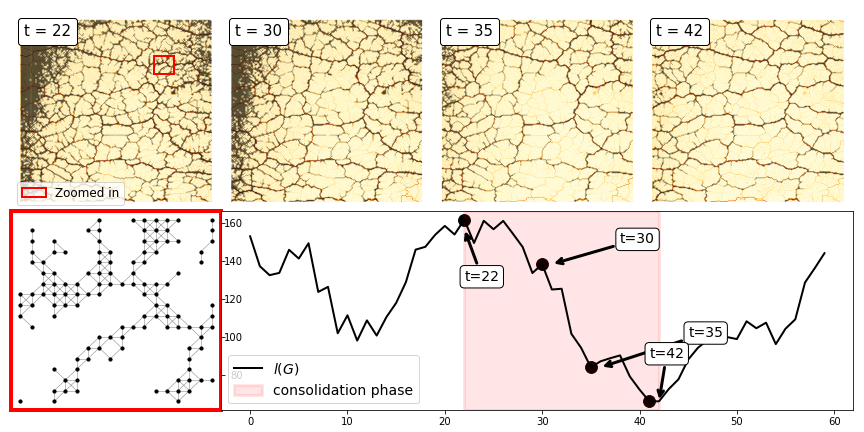}
\end{subfigure} 
\begin{subfigure}[b]{.8\textwidth}
\includegraphics[width=\textwidth]{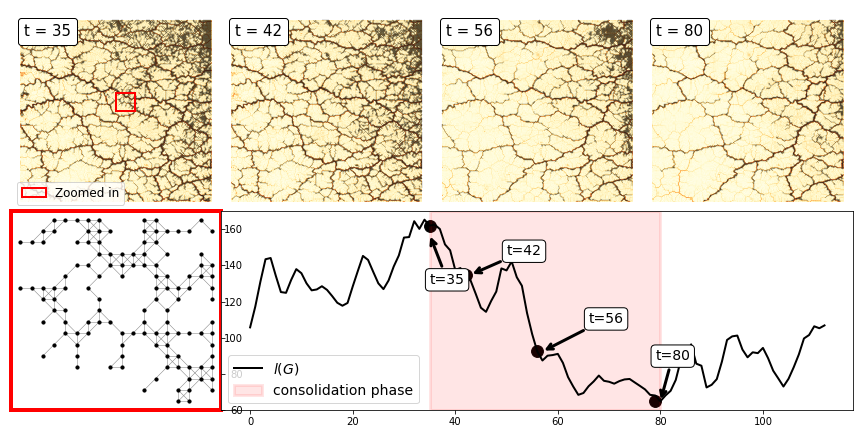}
\end{subfigure} 
\begin{subfigure}[b]{.8\textwidth}
\includegraphics[width=\textwidth]{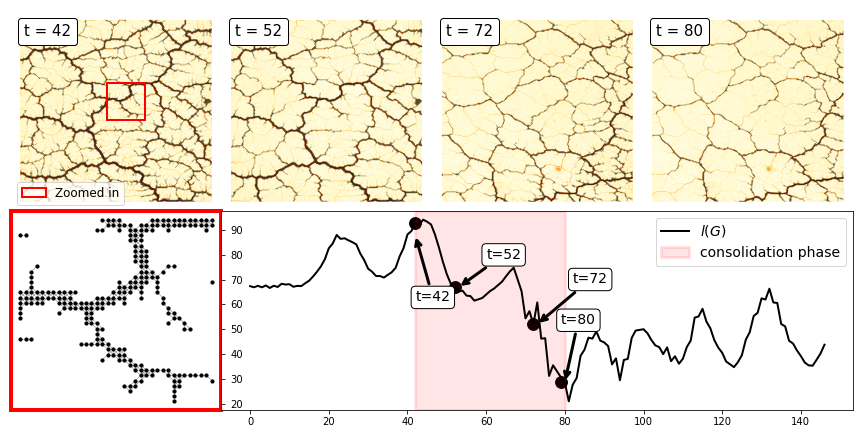}
\end{subfigure} 
\caption{\textbf{\textit{P. polycephalum} total length evolution}. From top to bottom: \textit{motion24, motion40} and \textit{motion79}. Plots are separated in couples. For every couple, the plots on top show both \textit{P. polycephalum} images and networks extracted from them. The network at the lower leftmost plot is a subsection of the graph shown inside the red rectangle on top. The plot at the bottom shows the total length as a function of time. The red shade in this plot highlights a tentative consolidation phase towards optimality.} \label{fig:image11}
\end{figure*}
 


\end{widetext}

\end{document}